\documentclass[aps,prb,showpacs,floatfix,superscriptaddress]{revtex4}
\usepackage{graphicx}

\begin{document}

\title{Intrinsic atomic scale modulations of the superconducting gap of {2H-NbSe$_2$}}

\author{I. Guillamon}
\affiliation{Laboratorio de Bajas Temperaturas, Departamento de F\'isica de la Materia Condensada \\ Instituto de Ciencia de Materiales Nicol\'as Cabrera, Facultad de Ciencias \\ Universidad Aut\'onoma de Madrid, 28049 Madrid, Spain}
\author{H. Suderow}
\affiliation{Laboratorio de Bajas Temperaturas, Departamento de F\'isica de la Materia Condensada \\ Instituto de Ciencia de Materiales Nicol\'as Cabrera, Facultad de Ciencias \\ Universidad Aut\'onoma de Madrid, E-28049 Madrid, Spain}
\author{F. Guinea}
\affiliation{Instituto de Ciencia de Materiales de Madrid,
Consejo Superior de Investigaciones Cient\'ificas,
 Campus de Cantoblanco, E-28049 Madrid, Spain.}
\author{S. Vieira}
\affiliation{Laboratorio de Bajas Temperaturas, Departamento de F\'isica de la Materia Condensada \\ Instituto de Ciencia de Materiales Nicol\'as Cabrera, Facultad de Ciencias \\ Universidad Aut\'onoma de Madrid, 28049 Madrid, Spain}

\begin{abstract}
We present scanning tunneling microscopy and spectroscopy measurements at 100mK in the superconducting material 2H-NbSe$_2$ that show well defined features in the superconducting density of states changing in a pattern closely following atomic periodicity. Our experiment demonstrates that the intrinsic superconducting density of states can show atomic size modulations, which reflect the reciprocal space structure of the superconducting gap. In particular we obtain that the superconducting gap of 2H-NbSe$_2$ has six fold modulated components at 0.75 mV and 1.2 mV. Moreover, we also find related atomic size modulations inside vortices, demonstrating that the much discussed star shape vortex structure produced by localized states inside the vortex cores, has a, hitherto undetected, superposed atomic size modulation. The tip substrate interaction in an anisotropic superconductor has been calculated, giving position dependent changes related to the observed gap anisotropy.
\end{abstract}

\pacs{74.25.Jb,74.50+r,74.70Ad} \date{\today}

\maketitle

\section{Introduction}

Today one of the more creative spaces in research related to superconductivity are atomic scale investigations by means of nanoprobes on well characterized surfaces of superconductors. Nanometric size inhomogeneities of the superconducting local density of states (LDOS) found using scanning tunneling microscopy and spectroscopy (STM/S) in high critical temperature superconductors, related to either the pseudogap state of the normal phase, or to scattering effects \cite{Pan00,Gomes07,Fischer07}, have clearly shed new light on the very nature of the pairing mechanism. However, it is generally considered that the superconducting LDOS is homogeneous at nanometric length scales, in ordered, metallic and pure samples \cite{Yazdani97,Hoffman02}. Here we report on the discovery of intrinsic atomic scale modulations of the superconducting gap using STM/S measurements at 100 mK.

We have found these modulations in one of the superconducting materials most commonly used in STM/S studies at low temperatures, the layered dichalchogenide superconductor 2H-NbSe$_{2}$. This compound has a critical temperature of T$_c$=7.1 K and develops a charge density wave (CDW) state below T$_{CDW}$=33 K. Historically, it has been one of the first superconducting materials where the superconducting LDOS has been studied in depth. This is not surprising, as it is easy to prepare flat and clean surfaces. Indeed, 2H-NbSe$_2$ is composed by layers of one sheet of Nb atoms between two sheets of Se atoms (Se-Nb-Se).  Adjacent layers are bonded with weak Van der Waals forces which leads to a pronounced two-dimensional character, making this material very easy to exfoliate. The last layer (surface layer) is formed by the Se atoms\cite{Coleman88,Dai93,Sacks98}. First observations of the vortex lattice with STM/S in a superconductor were made in this material\cite{Hess89,Hess90}. It is now believed that 2H-NbSe$_{2}$ is an important system where effects concerning interplay between charge order, superconductivity and a non trivial Fermi surface structure can be handled and studied in detail. The deviations from simple s-wave BCS theory of many experimental properties in this compound are notable and provide interesting analogies to superconductors close to so called quantum critical instabilities, and to the problem of multiband superconductivity\cite{Mathur98,Aoki01,Fletcher07,Valla04,Morosan06}. The Fermi surface consists of several nearly two dimensional concentric cylinders, derived from the Nb 4d orbitals and a single small three dimensional pancake like sheet, derived from the Se 4p orbitals \cite{Corcoran94,Johannes06}. The local superconducting density of states shows quasiparticle peaks far from the high anomalies expected from simple isotropic BCS theory \cite{Hess89,Hess90,Hess91}. This evidences the presence of a wide spread distribution of values of the superconducting gap, with gap sizes ranging between 0.6 meV and 1.4 meV, i.e. several hundreds $\mu$eV around the value expected from BCS theory (1.1 meV) \cite{Hess90,Rodrigo04c}. Further experiments highlight a close relationship between the band structure, the distribution of values of the superconducting gap, and the CDW state. The wide spread distribution of values of the superconducting gap have been proposed to reflect multi-band superconducting properties, with different superconducting gaps in the different Fermi surface sheets of this compound, in analogy to the intriguing two band superconductor MgB$_2$\cite{Rubio01,Liu01}. In particular, thermal conductivity and specific heat studies have shown the presence of at least two well differentiated values of the superconducting gap over the Fermi surface\cite{Boaknin03,Huang07}. Angular resolved photoemission experiments (ARPES) have been made only relatively close to T$_c$ (at 5.3 K) \cite{Yokoya01}, to find that Nb band gaps are around 1 meV. However, the Se band showed no gap opening, although London penetration depth studies give strong evidence for a relatively large gap in this band at low temperatures\cite{Fletcher07}. Moreover, de Haas van Alphen experiments in the mixed state and at very low temperatures detected a small superconducting gap of 0.6 meV, which was associated to the Se sheet\cite{Corcoran94}. Tunneling spectroscopy experiments demonstrated that the smaller size gap feature closes indeed rather fast (around 5.5 K) when increasing temperature \cite{Rodrigo04c}. On the other hand, most recent ARPES experiments \cite{Kiss07}, made with beautiful detail (although also at 5.3 K, which corresponds to state of the art experimental possibilities), have shown that the maxima of the superconducting and CDW gaps appear at the same position in reciprocal space, suggesting a cooperative interaction between charge fluctuations and superconductivity. This appears to go against the view obtained from pressure experiments, where it is clear that application of pressure decreases T$_{CDW}$, while T$_c$ increases\cite{Suderow05d}. However, T$_c$ continues increasing after the destruction of the CDW state, and the upper critical field shows also a peculiar and complex pressure dependence\cite{Suderow05d}, demonstrating that the pressure induced changes in the Fermi surface are non trivial and could produce in some unknown way a simultaneous destruction of CDW and an increase in T$_c$. On the other hand, the peculiar form of the internal electronic structure of vortices has also been put in connection with the ongoing debate about the anisotropic and/or multiband properties of 2H-NbSe$_2$\cite{Hess89,Hess90}. Experiments show that the LDOS in and around vortices has a characteristic star shape with a complex energy dependence, that has been explained in a great deal by taking into account the symmetry of the vortex lattice, the formation of quasiparticle bound states inside vortex cores, as well as a six-fold modulated superconducting gap\cite{Hayashi96}. However, the connection between this six-fold modulation and the zero field gap distribution, as revealed from the experiments, has not been made until now.

Within this debate, it is of importance to provide new precise information about the anisotropy and values of the superconducting gap at the lowest possible temperatures, i.e. close to the superconducting "ground state". Indeed, in a multiband superconductor the temperature dependence of the gaps found are expected to be non-trivial. Strong evidences for this has been obtained from the mentioned tunneling spectroscopy experiments, where changes in the distribution of values of the superconducting gaps in 2H-NbSe$_2$ have been followed as a function of temperature \cite{Rodrigo04c}. Moreover, it is equally important to relate the mixed phase observations with the zero field results. The atomic scale modulations of the superconducting gap reported here (found below 0.015T$_c$) show directly that the superconducting density of states has sixfold modulations at 0.75 mV and 1.2 mV, and therefore demonstrate that the associated gaps are six-fold modulated in reciprocal space. Measurements made under magnetic fields close to the vortex core demonstrate that these six fold atomic size modulations of the superconducting gap are superposed to the previously detected features of the LDOS inside the vortices.

\section{Experimental}

We use a very low temperature STM/S system installed in a dilution refrigerator. The set-up allows for in-situ clean preparation of tip and sample, following procedures described somewhere else \cite{Rodrigo04,Crespo06a,Guillamon07}. To ensure high resolution in voltage, all wires are filtered or attenuated. The experimental setup has been tested thoroughly before through the measurement of several low critical temperature superconductors such as Al (T$_c$ = 1.12 K), where we obtain clean BCS s-wave tunneling conductance curves with a negligible conductance at zero bias. The resolution in energy of our set-up lies below $15 \mu eV$ \cite{Suderow04,Rodrigo04,Crespo06a}. The atomic resolution topographic (STM) images are taken at constant tunneling current. The Se hexagonal lattice and the charge density wave are observed very clearly in 2H-NbSe$_2$ as in previous STM work on this compound (see e.g. \cite{Hess90,Sacks98,Guillamon07}). Full tunneling conductance vs. bias voltage $\sigma(V)$ curve ranging from voltages well above the superconducting gap are taken at each point of the topographic images, to construct STS images. The place where the $\sigma(V)$ curves are taken is uniquely identified with the atomic positions through comparison with the topography. The whole set-up is isolated from acoustic noise as well as vibrations from pumps and the building to allow for long term measurements. As a result, the conductance can be measured at each position with an accuracy of less than several per cent. In order to show the effects discussed below, we have to make a very close scan at atomic level. About a hundred tunneling conductance curves are taken for each group of three surface atoms, and full images cover regions including up to fifty $\times$ fifty atomic rows.

\section{Results and discussion}

\begin{figure}
\includegraphics[width=12cm,clip]{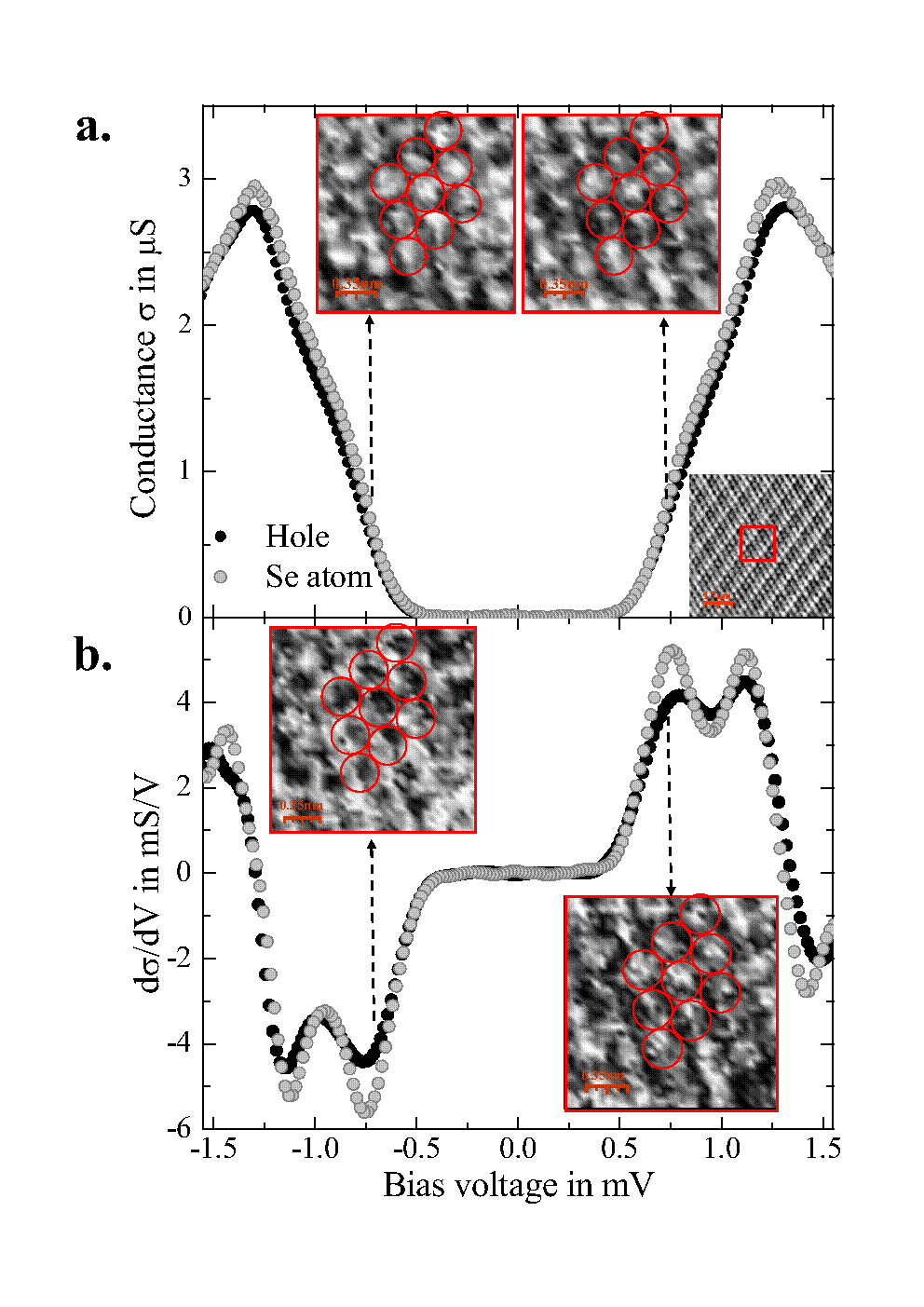}
\caption{(Color online) In the lower right inset of \textbf{a} we show typical topographic images obtained in 2H-NbSe$_2$, where the atomic Se lattice is clearly viewed, as well as the charge density wave, which produces an additional modulation in topography each three Se atoms. In the figure we show $\sigma(V)$ curves obtained on top of a Se atom and in between. STS images showing small groups of atoms constructed from these data at V$_1$ = $\pm$ 0.75 mV are shown in the top panels. In \textbf{b} we show $d\sigma/dV(V)$ at the same positions. The panels in the figure show STS images constructed from $d\sigma/dV$ at V$_1$. Full range in contrast in the STS panels from black to white represents, respectively a difference in $\sigma$ of 0.3$\mu$ S (\textbf{a}) and in $d\sigma/dV$ of 2mS/V (\textbf{b}). Circles in the images give the Se atomic positions as determined through topography. \label{fig1}}
\end{figure}

\begin{figure}
\includegraphics[angle=270,width=12cm,clip]{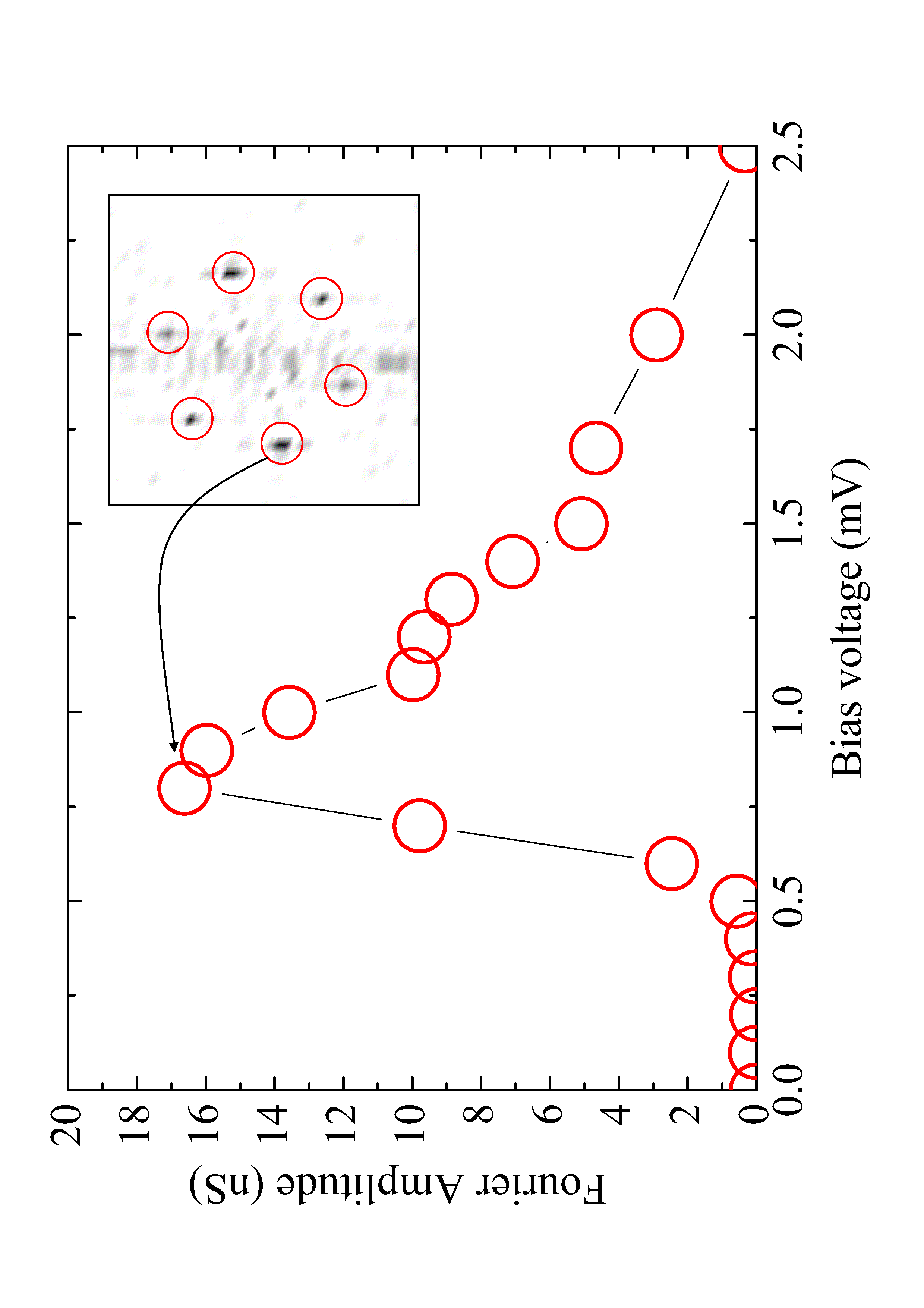}
\caption{(Color online) In the inset we show the Fourier transform of an STS image constructed from $\sigma(V= 0.75 mV)$ in 2H-NbSe$_2$ at 100 mK. Red circles represent the position of peaks in the Fourier transform of the corresponding topography image. The Fourier transform of STS images at other voltages is either flat, close to the Fermi level and at high voltages, or has the same form as the one shown. This is seen in the figure, where we represent the height of the six fold peaks of the Fourier transform shown in the inset as a function of the bias voltage (line is a guide to the eye). It is important to remark that no other peaks appear in the Fourier transform when changing the voltage, and that the features discussed in this paper fully disappear at voltages well above the superconducting gap.\label{fig2}}
\end{figure}

All $\sigma(V)$ curves are very similar to each other, and to the curves obtained in previous measurements \cite{Hess89,Hess91,Rodrigo04c}. The tunneling conductance smoothly increases at about 0.5 mV, has a small shoulder at about 0.75 mV and a rounded peak around 1.2 mV. However, here we are able to find slight, but relevant, differences between the curves taken at the top of the Se atoms, and those taken in between. In Fig.\ \ref{fig1} \textbf{a} we show tunneling conductance curves taken at these two positions. The shoulder at V$_1$ = $\pm$ 0.75 mV is better defined on $\sigma(V)$ taken over a Se atom. Correspondingly, STS images constructed from these curves at V$_1$ (top panels of Fig.\ \ref{fig1} \textbf{a}) clearly show a modulation of the conductance as a function of the position which reproduces the lattice of Se atoms. Note that this modulation corresponds to small changes in the tunneling conductance of the order of 10\%. The derivative of $\sigma(V)$ (Fig.\ \ref{fig1} \textbf{b}) has a peak around V$_1$ and another one at V$_2$ = $\pm$ 1.2 mV. STS images obtained from the derivative of $\sigma(V)$ at these voltages also clearly show the surface atomic lattice, as presented in the panels of Fig.\ \ref{fig1} \textbf{b}, where we show the result at V$_1$.

Actually, the full bias voltage dependence of the size of the conductance modulation is best discussed by analyzing the Fourier transform of the STS images. In the inset of Fig.\ \ref{fig2} we show the result of the Fourier transform of STS images obtained from the conductance at V$_1$. We observe a clear sixfold structure with Fourier peaks whose position coincides with the six fold peaks due to the surface Se lattice obtained in the Fourier transform of the associated STM topographic image. The height of these peaks strongly changes with bias voltage (Fig.\ \ref{fig2}). The six fold modulation appears around 0.6 mV, becomes best defined at V$_1$, with again a small feature at V$_2$, and disappears for higher voltages.

Note that, contrasting the results found in other superconducting systems, related to impurities and imperfections at or near the surface\cite{Fischer07}, the position of the Fourier peaks we find here does not change with voltage, i.e., there is no dispersion with energy. Indeed, periodic structures that have a significant dispersion in energy can be related to surface wave function scattering effects, which are much discussed in simple metals and in High T$_c$ superconductors, as they make it possible to obtain the electron dispersion relation by analyzing the Fourier transform of the STS images \cite{Pan00,Gomes07,Fischer07}. This is not the subject of present results, where we focus on measurements on neat surfaces, with a well ordered Se lattice. Here, no additional periodic structures, other than the atomic lattice, are resolved in the STS images at the voltage range studied here. These periodic structures are found only well within the quasiparticle peaks, and not at voltages above the superconducting gap, and show no changes of its reciprocal space position as a function of the bias voltage.

\begin{figure}
\includegraphics[angle=270,width=18cm,clip]{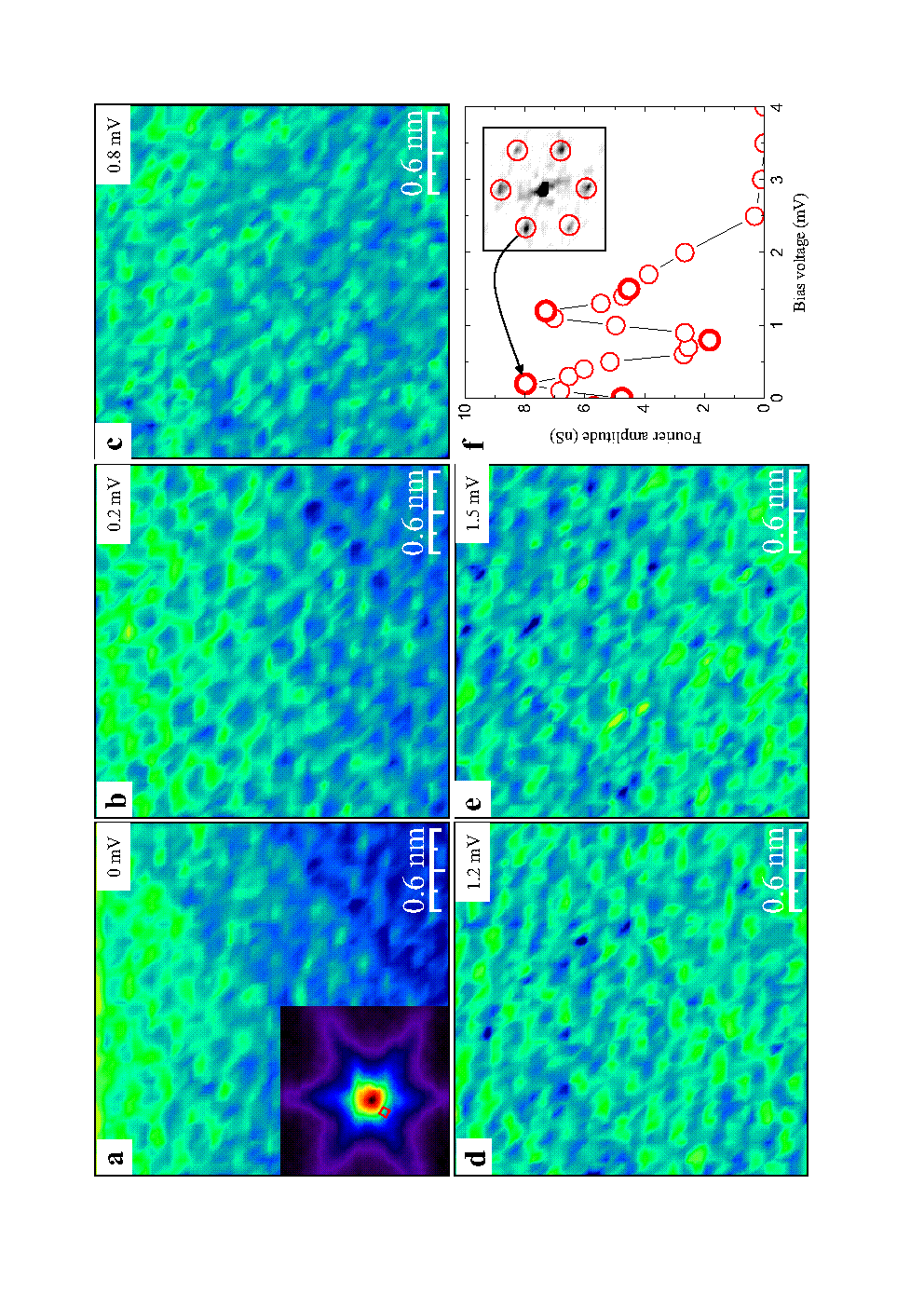}
\caption{In the inset of \textbf{a} (lower left corner) we show the STS image made at zero bias of a single vortex at 0.03 T. Color scale represents the value of the LDOS at the Fermi level, following previous representations of a vortex as measured with STS \protect\cite{Hess90,Guillamon07} (the whole color change corresponds to differences in the conductance from zero to about 2 $\mu$S). A zoom up of this image close to the vortex core, at the small red square of the inset, gives the STS images \textbf{a}-\textbf{e}, made at the bias voltages represented in each panel. Color scale in these panels, from green to blue, represents differences in the conductance of 0.2 $\mu$S, i.e. changes not higher than 10\% of the high bias conductance. The bias dependence of the height of the Fourier transform peaks of the images is shown in \textbf{f}. Points which have been obtained from the Fourier transform of images \textbf{a}-\textbf{e} are highlighted as greater circles. The inset shows the Fourier transform of the STS image at 0.2 mV, with red circles signalling the position of the peaks from the Fourier transform of the topography. Again, the effect of the bias voltage is to change the height of the peaks shown in the inset, becoming featureless at high bias voltages, as shown by the decrease of the Fourier amplitude in the figure. No new significantly remarked peaks appear when changing the voltage.\label{fig3}}
\end{figure}

In the mixed state we also find an atomic scale modulation of the superconducting features. It is well known that, close to the core of isolated vortices, the superconducting LDOS of 2H-NbSe$_2$ shows a complex position dependence at length scales comparable to the superconducting coherence length $\xi_0\approx$ 10 nm \cite{Hess89,Hess90}. The whole voltage dependence of the LDOS of the vortex lattice of 2H-NbSe$_2$, as found here in full agreement with Ref.\cite{Hess90}, is given in EPAPS Document No. []. Time scale reproduces the bias voltage of the STS experiment. The video begins at V$_{Bias}$ = - 2.5 mV and ends at V$_{Bias}$ + 2.5 mV, and each sequence corresponds to a change in bias voltage of 0.1 mV. The star shape shown in the inset of Fig.\ref{fig3} \textbf{a} is found close to the Fermi level, which appears at the middle of the video. Indeed, the way the gap closes inside a vortex core is not trivial, and one of the most characteristic features is the six fold star shape structure which is observed on STS images obtained from the zero bias conductance. In the inset of Fig.\ \ref{fig3} \textbf{a} we show the star shape of a single vortex at 0.03 T (see also \cite{Hayashi96,Nishimori04}).

Close to the center of the vortex, at the small square represented in the inset of Fig.\ \ref{fig3} \textbf{a}, STS images made at atomic length scales again show the atomic surface lattice, only at voltages where superconducting features appear on the LDOS. In Figs.\ \ref{fig3} \textbf{a}-\textbf{e} we show STS images at different bias voltages. Note that the zero bias voltage image shows a smooth color change from more blue at the bottom to more green at the top. There is a smooth increase in the conductance, superposed to the atomic size modulation, which reproduces the smooth variation in the tunneling conductance that leads to the star shape shown in the inset of Fig.\ref{fig3} \textbf{a}. This variation is less pronounced at higher bias voltages, although it is always present and superposed to the atomic size modulations discovered here (Figs.\ \ref{fig3} \textbf{a}-\textbf{e}). Note also that the star shape structure is aligned with the hexagonal modulation of the superconducting gap and with the atomic lattice. The bias voltage dependence of the height of the corresponding Fourier peaks (Fig.\ \ref{fig3} \textbf{f}) is very different than the one found at zero field. Although in both cases no modulation is observed at voltages well above the quasiparticle peaks, the modulation under magnetic fields is observed also close to the Fermi level, as the superconducting LDOS is not zero close to the vortex core. A slight super-modulation at the reciprocal wavevectors of the charge density wave is also observed in some of these images. The height of the corresponding Fourier peaks are much smaller than the ones shown in Fig.\ \ref{fig3} \textbf{f}, and its energy dependence is very difficult to extract from present data, although they appear to be better defined close to the Fermi level. Higher resolution experiments may be able to resolve these details in future.

It is important to remark that the atomic surface lattice is viewed in the STS images only at voltages of the order of the position of the superconducting features in the LDOS. For voltages well above the highest value of the superconducting gap, the STS images are featureless, as seen in the strong decrease in the height of the Fourier peaks, shown in Fig.\ \ref{fig2} and in Fig.\ \ref{fig3} \textbf{f}. This rules out any effects related to tip-substrate coupling phenomena not intrinsically due to the local superconducting properties.

It is well known that the atomic scale LDOS of the sample, as viewed in the STS experiment, is formed by contributions from different parts of the Fermi surface, weighted by details of the interaction between tip and sample \cite{Fischer07}. Many approaches are being developed and applied to understand spectroscopic imaging in different kinds of metallic or semiconducting surfaces. The superconducting state has received less attention, probably because it is generally considered that, in a first approximation and in pure systems, the superconducting gap is relatively homogeneous over distances of the order of $\xi_0$ \cite{Yazdani97,Hoffman02,Balatsky06}. Nevertheless, it is clear that non trivial variations of the superconducting density of states in reciprocal space, as found in multiband superconductors or in systems with a wide distribution of values of the superconducting gap around the Fermi surface, have to appear as modulated structures in the real space LDOS. Therefore, the modulated features observed on the STS images at V$_1$ and V$_2$ show the presence of concomitant modulations of the superconducting gap on reciprocal space. The six fold modulations at V$_1$ = 0.75 mV and V$_2$ = 1.2 mV observed in our experiments are associated to sixfold reciprocal space modulations of the superconducting gap of 2H-NbSe$_2$.

The Fermi surface position where this six fold modulated gap structure appears is more difficult to establish independently from our experiment at present. In the model described below, we give a first approach to extract more detailed reciprocal space information of the superconducting gap from its atomic size modulations. It is however useful to first compare our result with previous experiments. For instance, while the size of the distribution of values of the superconducting gap has been long known, since the seminal tunneling experiments of Hess et al.\cite{Hess89,Hess90} and subsequent measurements \cite{Boaknin03,Rodrigo04c,Fletcher07,Huang07}, the discussion about the reciprocal space dependence has been limited to the values of the different gaps in different sheets of the Fermi surface. Eventual anisotropies and its symmetry remained difficult to extract directly from the experimental data available. For example, the connection of the gap anisotropy and the six fold star shape structure of the vortex cores, found in Ref.\cite{Hess90} was made through the ad-hoc introduction of a modulated superconducting gap in theoretical calculations\cite{Hayashi96}.

On the other hand, very recent ARPES experiments have shown that the gap has maxima at points in reciprocal space that are directly connected with the CDW vector\cite{Kiss07}. A direct comparison to our experiment is difficult, because ARPES measurements were made, although with great detail, at temperatures very close to T$_c$ (at 5.3 K, while T$_c$=7.1 K), and the fits to the spectra obtained needed the addition of a smearing parameter to get the values of the superconducting gap. Our data, by contrast, are taken at 0.1 K, and have a flat and zero conductance at the Fermi level, giving neat spectra that do not show any sign of smearing effects. The ARPES experiment shows that there are six-fold modulations of the superconducting gap along directions that coincide with the atomic lattice and CDW reciprocal wavevectors. Some of the corresponding gap magnitudes are very similar to the ones found here (e.g. 1.2 mV and 1 mV along the $\Gamma$-$K$ symmetry line\cite{Kiss07}). Indeed, the anisotropy observed in the present experiment is very close to the one reported in some of the ARPES results. Although the CDW wave-vectors are practically not observed in the Fourier analysis of the atomic size modulations of the superconducting gap shown here (except in the experiment shown in Figs.\ \ref{fig3}, where the corresponding signal is very weak), the directions of the sixfold anisotropy observed here is the same as the CDW wave vector direction.

Additional angular changes of the superconducting gap are also found in ARPES data for significantly lower values of the gap, going down to 0.3 mV. This is fully absent in our experiment, as the LDOS, at zero field, is flat and equal to zero (within one per cent of the high voltage value) below 0.5 mV. However, as mentioned previously, the tunneling experiments of Ref.\cite{Rodrigo04c} demonstrate that the smaller sized gaps of 2H-NbSe$_2$ close when approaching T$_c$, so that the apparent discrepancy is clearly due to temperature differences. Note that therefore, a significantly sized anisotropy in the Se sheets at low temperatures, which may have an additional influence to the results found here, should not be fully excluded at present.

Anyhow, our experiment gives a determination of the values of the six-fold anisotropy in the superconducting gap with unprecedented detail, free from temperature and smearing effects, and also additional local information close to the vortex cores. Indeed, the atomic size modulations found within the vortex core should come as a consequence, within these considerations, of the closing of the superconducting gap, and produce the changes in the height of the Fourier peaks towards energy values closer to the Fermi level. These changes, however, are largely non trivial, as seen from Fig.\ \ref{fig3} \textbf{f}. First, the Fourier peaks shown in Fig.\ \ref{fig3} \textbf{f} reveal that the smaller sized gaps tend to close earlier when approaching the core, so that the anisotropic gap at V$_2$ remains stable pretty close to the core, and the anisotropic gap at V$_1$ shifts to the Fermi level. Indeed, the Fourier peak at V$_2$ remains practically at the same position as in zero field. And second, the sixfold modulation of the superconducting gaps is maintained in its full importance well within the vortex core, as evidenced by the height of the Fourier peaks, which are of the same order at V$_2$ than close to the Fermi level, e.g. at V$_1$ (Fig.\ \ref{fig3} \textbf{f}). So the effect of the magnetic field inside the core is not a simple smearing effect, and the much more tricky formation of localized states, combined with gap anisotropies, as introduced in Ref.\cite{Hayashi96}, must be considered as a central issue in the interpretation of vortex lattice images using STS.

\section{Theory}

The development of a microscopic model of the dependence of the observation of superconducting properties as a function of atomic scale tip substrate interactions, taking precisely into account the whole Fermi surface of 2H-NbSe$_2$ is out of the scope of the present paper. Nevertheless, we have verified that a simple description of the band structure and tip substrate interaction in 2H-NbSe$_2$ leads to intrinsic modulations of the superconducting density of states on atomic length scales, related to the Fermi surface dependence of the superconducting properties. Within this description, the anisotropies of the order parameter lead to modulations in the density of states measured by a local probe on length scales comparable to the lattice spacing. The two ingredients of the model which lead to changes in the superconducting state are the degeneracies of the electronic bands at high symmetry points in the Brillouin Zone, and a modulation of the superconducting gap, compatible with the lattice and with previous proposals to explain some features of 2H-NbSe$_2$.

First, we present the model Hamiltonian we use. We approximate the band structure by a nearest neighbor tight binding model which includes the $d_{x^2-y^2}$ and $d_{xy}$ orbitals. We consider a single hexagonal Nb layer. This model is an approximation to a more complex three dimensional model which includes Se $p_z$ orbitals and Nb $s$ and the other Nb $d$ orbitals. If the interlayer couplings are weak, and the Se and Nb orbitals not
included initially in the model are sufficiently far from the Fermi surface, they can be included perturbatively. The resulting
hamiltonian is given by a $2 \times 2$ matrix, with the same degeneracies and symmetries as one defined solely in terms of  the
$d_{x^2-y^2}$ and $d_{xy}$ orbitals.

The model includes two nearest neighbor hoppings, $t_{dd \sigma}$ and $t_{dd  \pi}$.
We define:
\begin{eqnarray}
c_1 &= &\cos ( k_x a ) \nonumber \\
c_2 &= &\cos \left( \frac{k_x a}{2} + \frac{\sqrt{3} k_y a}{2}
      \right)  \nonumber \\
c_3 &= &\cos \left( - \frac{k_x a}{2} + \frac{\sqrt{3} k_y a}{2}
      \right)
\end{eqnarray}
where $a$ is the lattice constant. The hamiltonian is:
\begin{equation}
{\cal H} \equiv \left( \begin{array}{cc} t_{dd \sigma} \left( 2 c_1
+
      \frac{c_2 + c_3}{2} \right) + \frac{3 t_{dd \pi} ( c_2 + c_3 )}{2} &
      \frac{\sqrt{3} (  t_{dd \sigma} - t_{dd \pi} ) ( c_2 - c_3 )}{2} \\
 \frac{\sqrt{3} (  t_{dd \sigma} - t_{dd \pi} ) ( c_2 - c_3 )}{2} &
       \frac{3 t_{dd \sigma} ( c_2 + c_3 )}{2} +   t_{dd \pi} \left( 2 c_1 +
      \frac{c_2 + c_3}{2} \right)
      \end{array} \right)
\label{hamil}
\end{equation}
We can expand the hamiltonian around $k_x , k_y = 0$:
\begin{equation}
{\cal H}_\Gamma \equiv \left( \begin{array}{cc} 3 ( t_{dd
 \sigma} +
      t_{dd \pi}
      ) \left[ 1 + \frac{( k_x^2 + k_y^2 ) a^2}{4} \right] + \frac{3 ( t_{dd \sigma} - t_{dd \pi} ) (
      k_x^2 - k_y^2 ) a^2}{8} & \frac{3 ( t_{dd \sigma} - t_{dd \pi} ) k_x
      k_y a^2}{4} \\  \frac{3 ( t_{dd \sigma} - t_{dd \pi} ) k_x
      k_y a^2}{4} &   3 ( t_{dd
\sigma} +
      t_{dd \pi}
      ) \left[ 1 + \frac{( k_x^2 + k_y^2 ) a^2}{4} \right] - \frac{3 ( t_{dd
          \sigma} - t_{dd \pi} ) (
      k_x^2 - k_y^2 ) a^2}{8} \end{array} \right)
\end{equation}
which can also be written in angular coordinates as:
\begin{equation}
{\cal H}_\Gamma \equiv  3  ( t_{dd \sigma} +
      t_{dd \pi}
      ) \left( 1 + \frac{k^2 a^2}{4}  \right) {\cal I} + \frac{3 ( t_{dd \sigma} - t_{dd \pi} )
      k^2 a^2 \cos (
      2 \theta )}{8} \sigma_z + \frac{3 ( t_{dd \sigma} - t_{dd \pi} ) k^2
      a^2 \cos (
      2 \theta )}{8} \sigma_x
\end{equation}
where ${\cal I}$ is the $( 2 \times 2 )$ unit matrix, and $\sigma_z$
and $\sigma_x$ are Pauli matrices. The eigenvalues of this matrix are
\begin{equation}
\epsilon_k \approx 3 ( t_{dd \sigma} + t_{dd \pi} ) (1 + \frac{k^2
a^2}{4}) \pm 3 ( t_{dd \sigma} -
      t_{dd \pi} ) ( k a )^2  / 8
\end{equation}
and the eigenfunctions are:
\begin{eqnarray}
\left| \Psi_{\vec{k}} \right\rangle &\equiv &\cos ( \theta^\Gamma )  \left| dd\pi
      \right\rangle_{\vec{k}} + \sin ( \theta^\Gamma ) \left| dd\sigma \right\rangle_{\vec{k}}
      \nonumber \\
\left| \Psi_{\vec{k}} \right\rangle &\equiv &-\sin ( \theta^\Gamma )  \left| dd\pi
      \right\rangle_{\vec{k}} + \cos ( \theta^\Gamma ) \left| dd\sigma
      \right\rangle_{\vec{k}}
\label{vw_G}
\end{eqnarray}
where $\vec{k}$ is a vector at the Fermi surface, $\theta^\Gamma$ is the angle
 which denotes the position of $\vec{k}$ ar the Fermi surface and $| dd\pi
 \rangle_{\vec{k}}$ nd $| dd\sigma\rangle_{\vec{k}}$ are Bloch waves built
 from $dd\pi$ and $dd\sigma$ atomic orbitals.
 The bands
      are isotropic, and the eigenvectors rotate
      around the $\Gamma$ point, with a winding number of $2 \pi$.

A similar expansion can be done around the $K$ and $K'$ points. We define, for the $K$ point:
\begin{eqnarray}
k_x a &= &\frac{\pi}{3} + \delta k_x a \nonumber \\
k_y a &= &\frac{\sqrt{3} \pi}{3} + \delta k_y a
\end{eqnarray}
The hamiltonian becomes:
\begin{equation}
{\cal H}_K \equiv \left( \begin{array}{cc} - \frac{3}{2} ( t_{dd \sigma}  +
    t_{dd \pi} ) +
    \frac{3 \sqrt{3}}{4} ( t_{dd \sigma} - t_{dd \pi} )
    \delta k_x a &  \frac{3 \sqrt{3}}{4} ( t_{dd \sigma} -
t_{dd \pi} )  \delta k_y a  \\  \frac{3 \sqrt{3}}{4} ( t_{dd \sigma} -
t_{dd \pi} )  \delta k_y a  &- \frac{3}{2} ( t_{dd \sigma}  +
    t_{dd \pi} ) -
    \frac{3 \sqrt{3}}{4} ( t_{dd \sigma} - t_{dd \pi} )  \delta k_x a
    \end{array} \right)
\end{equation}
or, in angular coordinates:
\begin{equation}
{\cal H}_K \equiv - \frac{3}{2} ( t_{dd \sigma} + t_{dd \pi} ) {\cal
I} + \frac{3 \sqrt{3}}{4} ( t_{dd \sigma} - t_{dd \pi} ) \delta k a
\cos ( \theta^K ) \sigma_z + \frac{3 \sqrt{3}}{4} ( t_{dd \sigma} -
t_{dd \pi} ) \delta k a \sin ( \theta^K ) \sigma_x
\end{equation}
with eigenstates
\begin{equation}
\epsilon_k = - 3 ( t_{dd \sigma} + t_{dd \pi} ) /
2 + 3 \sqrt{3}  ( t_{dd \sigma} - t_{dd \pi} ) | \delta k a | / 4
\end{equation}
and eigenfunctions:
\begin{eqnarray}
\left| \Psi_{\vec{k}} \right\rangle &\equiv &\cos \left( \frac{\theta^K}{2} \right)  \left| dd\pi
      \right\rangle_{\vec{k}} + \sin \left( \frac{\theta^K}{2} \right)  \left| dd\sigma \right\rangle_{\vec{k}}
      \nonumber \\
\left| \Psi_{\vec{k}} \right\rangle &\equiv &-\sin \left( \frac{\theta^K}{2} \right)  \left| dd\pi
      \right\rangle_{\vec{k}} + \cos \left( \frac{\theta^K}{2} \right) \left|
      dd\sigma \right\rangle_{\vec{k}}
\label{wv_K}
\end{eqnarray}
 The winding number in this case is $\pi$.

\begin{figure}
\begin{center}
\includegraphics[angle=270,width=16cm]{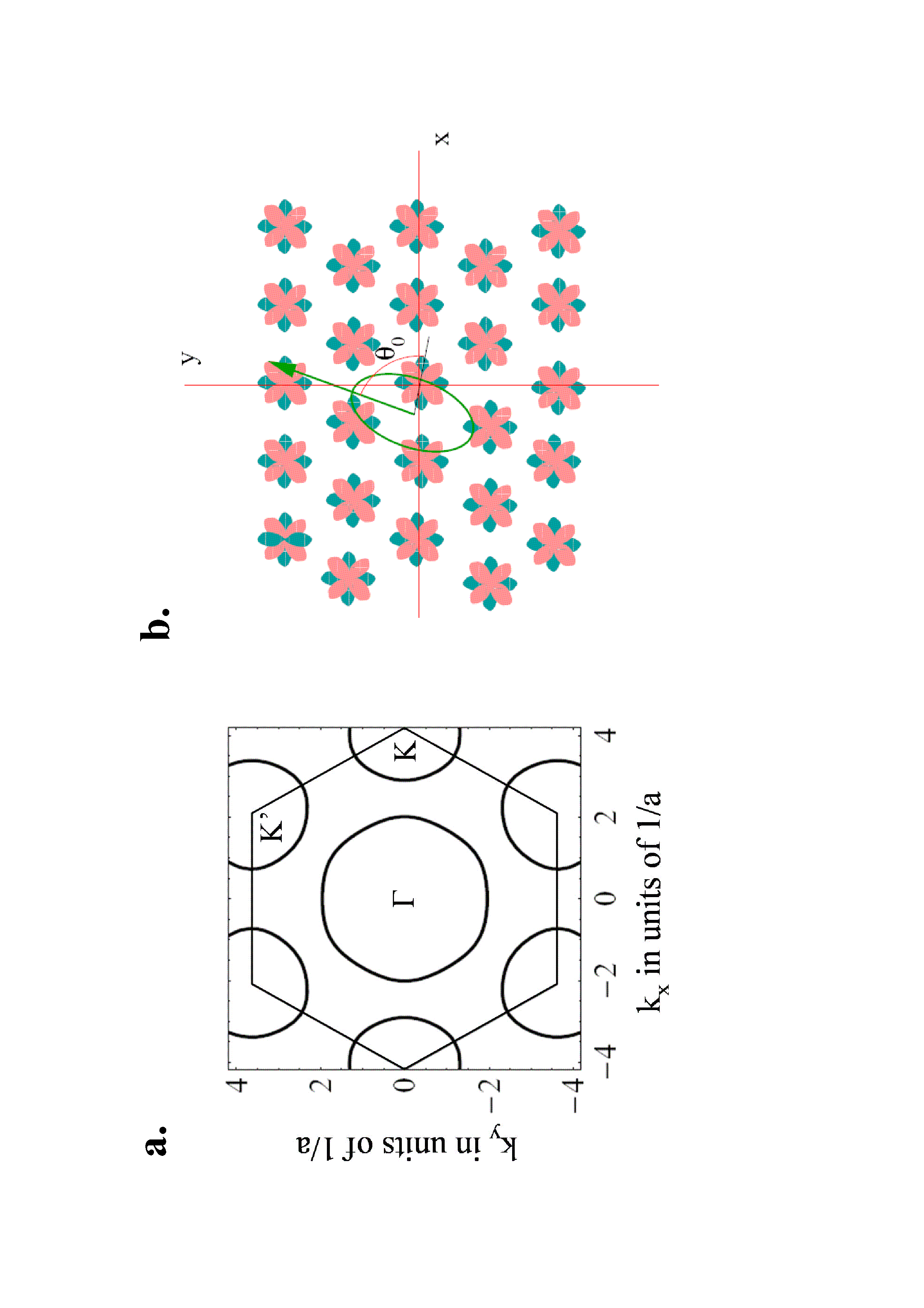}
\end{center}
\caption{ In a. we show Fermi surface obtained from the model discussed here (a is the lattice spacing). High symmetry points are labeled as $\Gamma$, $K$ and $K'$. In b. we represent schematically the anisotropy of the tip orbitals. The ellipse in the figure graphically shows the anisotropy of the tip-substrate interaction. The Nb $d_{x^2-y^2}$ and $d_{xy}$ orbitals are represented as in plane lobes (green and red respectively). $\theta_0$ is the angle between the long axis of the anisotropy of the tip and the straightest line connecting the center of the tip to the nearest Nb atom,
schematically represented here at the center of the 2D real space coordinate system.} \label{fig4}
\end{figure}

The above analysis justifies the almost circular Fermi surface sheets obtained from diagonalizing the hamiltonian, eq.[\ref{hamil}], shown in Fig.\ \ref{fig4} \textbf{a}. The symmetries of the eigenfunctions are also consistent with the general behavior of a Hamiltonian with twofold degeneracy defined in
an hexagonal lattice. In particular, the linear (Dirac) dispersion found near the $K$ and $K'$ points is analogous to that observed at the corners of the Brillouin Zone in graphene \cite{Geim07,Manes07}.

The model can be enlarged by including the $d_{z^2}$ orbitals of Nb and the $p_z$ orbitals of Se, and also interlayer hoppings, leading to four
bands which disperse in all directions. For reasonable parameters, however, the results do not change qualitatively with respect to the simpler two band
model described here.

We now consider the tip-substrate interaction. The current through the tip measures the LDOS. If the tip is not placed on top of a Nb site, it will couple more strongly to a certain combination of $d_{x^2-y^2}$ and $d_{xy}$ orbitals. We consider a tip which interacts with more than one atom. The interaction is assumed to be anisotropic. In Fig. \ref{fig4} \textbf{b} we show schematically this anisotropy through an ellipse centered at the position of the tip. We assume that the tunneling current is determined by a single combination of atomic orbitals at the tip, $|tip \rangle $. The current will be proportional to the overlap of this combination of orbitals and the atomic orbitals at the substrate. These overlaps can be written in terms of $\sigma$ and $\pi$ components.

\begin{figure}
\includegraphics[angle=270,width=12cm,clip]{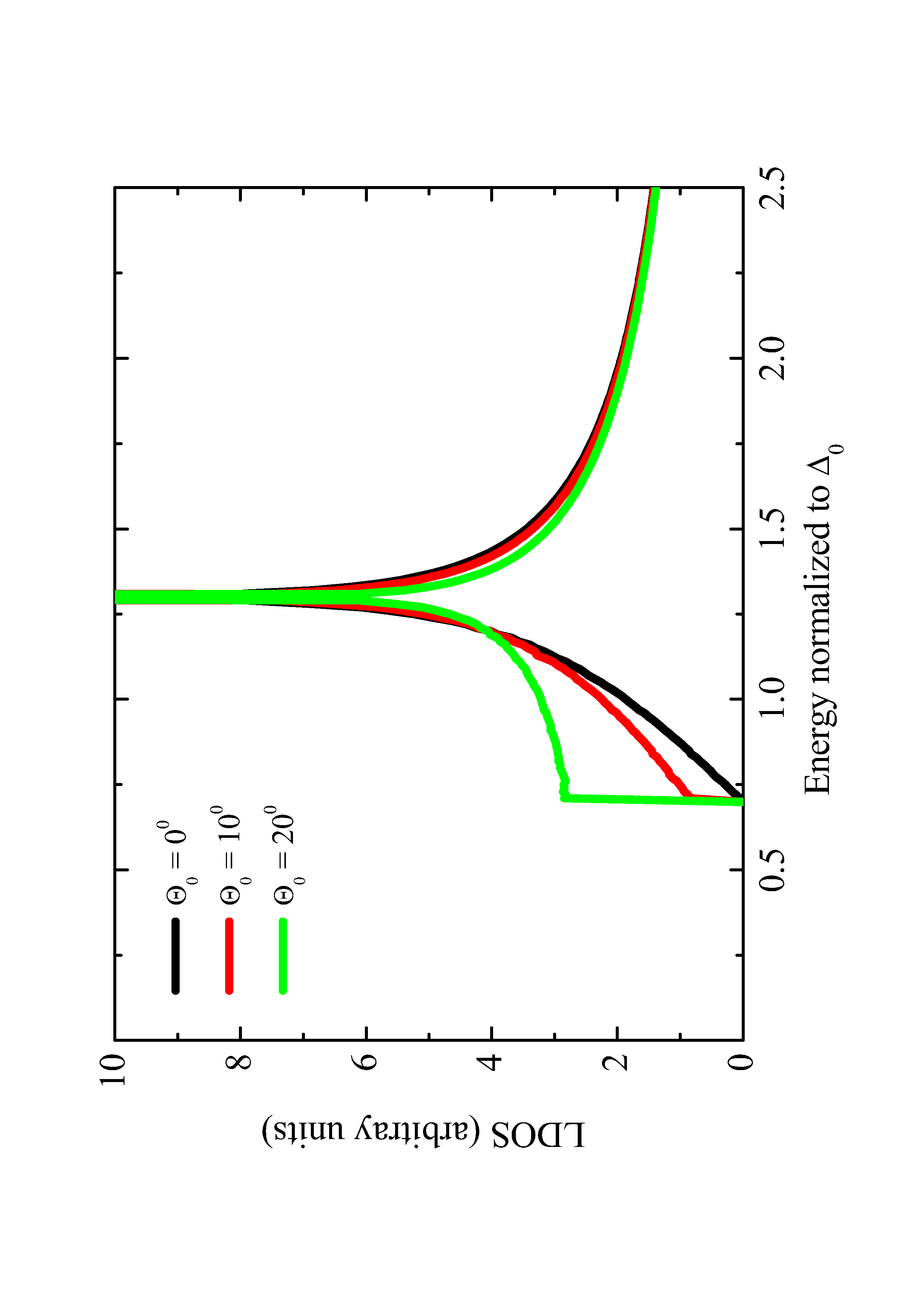}
\caption{(Color online) In the figure we plot the LDOS for different angles when the tip describes a circle around a Nb atom ($\theta_0 = 0^\circ$, black, $10^\circ$ in red; and $20^\circ$ in green). We show the result near the $K$ point. The parameters used for the superconducting gap are $\Delta_0 = 1$ meV and $\Delta_1 = 0.3$ meV. \label{fig5}}
\end{figure}

The relevant orbitals in the substrate are Bloch waves near the Fermi energy,
derived from the atomic orbitals as discussed previously. We
consider the Fermi surface sheet near the $K$ point, and the Bloch waves $|
d_{x^2-y^2}  \rangle_{\vec{k}}$ and $|d_{xy} \rangle_{\vec{k}}$, as defined
in eq.(\ref{wv_K}). A STM tip, at some
distance of the Nb atoms, will couple to a combination of the $d_{x^2-y^2}$
and $d_{xy}$ orbitals, whose in plane orientation with respect to the lattice
axis can be described by $\theta_0$. $\theta_0$ is the angle between the long
axis describing the anisotropy of the tip orbitals (long axis of the
schematic ellipse shown in Fig. \ref{fig4} \textbf{a}), and the line between the tip
and the nearest located atom. Then, the overlaps are, for the $\sigma$
contribution:

\begin{equation}
\begin{array}{lll}
\langle tip | d_{x^2-y^2} \rangle_{\vec{k}} &\propto \sum_{i,j} e^{i
\delta k
  r_{ij} \cos ( \theta_{ij} - \theta^K )} f ( r_{ij}
) \cos [ 2 ( \theta_{ij} - \theta_0  ) ] &\approx \\ &\approx \int e^{i
\delta k r \cos (
  \theta - \theta^K  )} f ( r ) \cos [ 2 (
\theta - \theta_0  ) ] r dr d \theta &= \cos [ 2 ( \theta^K - \theta_0  )
 ] \int
J_2 ( \delta k r ) f ( r ) r d r\\
\langle tip | d_{xy} \rangle_{\vec{k}} &\propto \sum_{i,j} e^{i \delta k
r_{ij} \cos (  \theta_{ij} - \theta^K  )} f (
 r_{ij} ) \sin [ 2 (
\theta_{ij} - \theta_0  ) ] &\approx \\ &\approx \int  e^{i \delta k r
\sin (
  \theta - \theta^K  )} f ( r ) \sin [ 2 (
\theta - \theta_0   ) ] r dr d \theta &= \sin [ 2  ( \theta^K - \theta_0
) ] \int J_2 ( \delta k r ) f ( r ) r d r \label{overlap}
\end{array}
\end{equation}

where the labels $i,j$ stand for lattice positions, defined with respect to the axis shown in Fig.\ \ref{fig5}, and the angular coordinates $r_{ij}$ , $\theta_{ij}$ are used. There is a radial modulation of the contribution from each lattice site, included in the function $f ( r_{ij} )$. The function $J_2 ( x )$ is a Bessel function. $\theta^K$ is used for the angular coordinate in reciprocal space with respect to the $K$ point, as previously. A similar expression is found for the $\pi$ contribution, except that the factors $\cos [ 2 (\theta^K - \theta_0 ) ]$ and $\sin [ 2  ( \theta^K - \theta_0 ) ]$ are interchanged.

The anisotropic contributions in eq.(\ref{overlap}) have to be integrated
over the Fermi surface. As a further ingredient, we need to give the simplest
non isotropic modulation of the superconducting gap, compatible with the
hexagonal symmetry of the lattice (extended s-wave with a finite
non-isotropic component, see \cite{Hayashi96,CastroNeto01,Suderow05d,Fletcher07}), which
can be written as $\Delta_0^K + \Delta_1^K \cos ( 3 \theta^K )$ around the
$K$ point. Then, the density of states measured by the tip includes a
position dependent term:

\begin{equation}
\delta G_{tip,K} ( \omega ) = \int d \theta^K \frac{\left| \cos [
2 ( \theta^K -
    \theta_0 ) ] \cos \left( \frac{\theta^K}{2} \right) + \sin [ 2
   (  \theta^K - \theta_0  ) ] \sin \left( \frac{\theta^K}{2} \right)
 \right|^2
    \omega}{\sqrt{\omega^2 - [ \Delta_0^K + \Delta_1^K \cos ( 3 \theta^K )
    ]^2}}
\label{dos_K}
\end{equation}
A similar equation describes the contribution from the $K'$ point, with
$\Delta_1^{K'} = - \Delta_1^K$. The hexagonal symmetry is recovered from
adding both terms.

The same analysis, when applied to the states around $\Gamma$ leads to the correction to the Green's function:

\begin{equation}
\delta G_{tip,\Gamma} ( \omega ) = \int d \theta^\Gamma
\frac{\left| \cos [ 2 (
    \theta^\Gamma -\theta_0  ) ] \cos ( \theta^\Gamma ) + \sin [ 2
 (   \theta^\Gamma - \theta_0   ) ] \sin ( \theta^\Gamma ) \right|^2
    \omega}{\sqrt{\omega^2 - [ \Delta_0^\Gamma + \Delta_1^\Gamma \cos ( 6
    \theta^\Gamma ) ]^2}}
\label{dos_G}
\end{equation}

The expression for $\delta G_{tip,K}$ in eq.(\ref{dos_K}) leads to a significant dependence on
$\theta_0$, as the numerator when expanded in angular harmonics, includes
components proportional to $3 \theta^K$, as in the denominator. This is not
the case for the contribution from the Fermi surface near the $\Gamma$ point,
eq.(\ref{dos_G}).
 Hence, the effects obtained within this approach
are clearly associated to the proximity to the edge of the Brillouin zone.

In Fig.\ \ref{fig5} we present the contribution to the LDOS calculated by allowing the tip to describe a circular path around a Nb atom, as a function of the angle $\theta_0$. There are clear position dependent changes, produced by the degeneracy of the Nb orbitals, combined with the modulation of the superconducting gap.

The changes in the electronic structure, as measured by the STM tip, near a vortex can be obtained by solving the Eliashberg equations in a vector potential. They require a knowledge of the modulation of the gap, which will be reflected in the density of states. Qualitatively, and using the simpler Ginzburg-Landau equations, it can be expected that the non trivial hybridization around the Fermi surface will lead to changes in the gradient term in the
Ginzburg-Landau expansion. This term modifies the superconducting gap when the superconducting order parameter is modulated spatially over distances comparable to the coherence length, as it happens near a vortex. The gap is reduced. A proportional reduction of $\Delta_0$ and $\Delta_1$ near the core of a vortex leads to changes in the density of states in qualitative agreement with the experiments.

\section{Conclusions}

In summary, we have observed a sizable modulation of the superconducting density of states at interatomic distances, localized at energies corresponding to the size of the different superconducting gaps found over the Fermi surface of 2H-NbSe$_2$, and at reciprocal space vectors fully coinciding with the atomic Se surface lattice. Moreover, the star shape vortex structure found by Hess et al. \cite{Hess90}, also shows a significant atomic size modulation, not detected previously, close to the vortex core. These modulations appear again localized at the lattice reciprocal space, with a small contribution at spacings corresponding to the CDW, and at energies that are however closer to the Fermi level. We have performed a first simple analysis that shows that a combination of the degeneracies of the wavefunctions near high symmetry points, and a finite angular dependence of the superconducting gap around the Fermi surface, are enough to induce an angular modulation on the current measured by an STM tip, only at voltages within the superconducting gap. Other approaches involving different symmetries for the tip substrate coupling, or different superconducting gap modulations can also lead to a spatially varying local superconducting density of states. A more sophisticated microscopic model taking into account more precisely the band structure of 2H-NbSe$_2$ and possible tip substrate coupling functions should lead, through the comparison to the experiment, to new, additional insight into the properties of the superconducting phase of 2H-NbSe$_2$. From the present approach, it already becomes clear that the atomic size features in the STS curves open a new route to measure the reciprocal space structure of the superconducting gap, not only in 2H-NbSe$_2$, but also in other systems with a significant gap anisotropy. In the present work we have identified the two energy scales (V$_1$ = 0.75 mV and V$_2$ = 1.2 mV) corresponding to the two main gap magnitudes open over the Fermi surface in this compound, as well as a six fold modulation associated with them. We also give first evidence for an additional atomic size modulation of the superconducting gap at the vortex cores of 2H-NbSe$_2$.

The present experiment breaks with the conventionally accepted view that the intrinsic superconducting properties are homogeneous at length scales below the superconducting coherence length. The atomic scale structures found at zero field and within vortices are a surprising feature of superconductivity, which is possibly unique among macroscopic quantum ordered states.

\section{Acknowledgments}

We are indebted to P. Rodi\`ere  for providing us with samples and to J.G. Rodrigo, J.P. Brison, E. Bascones and V. Crespo for discussions. We also acknowledge help of A.I. Buzdin, A.P. Levanyuk, K. Behnia, P.C. Canfield, M. Johannes and J. Flouquet. The Laboratorio de Bajas Temperaturas is associated to the ICMM of the CSIC. This work was supported by the Spanish MEC (CONSOLIDER INGENIO 2010 program and FIS2004-02897), by the Comunidad de Madrid through program "Science and Technology in the Millikelvin" (S-0505/ESP/0337), and by NES and ECOM programs of the ESF.


\end{document}